\documentclass{mem}
\usepackage{natbib}
\usepackage{graphicx}
\usepackage[a4paper]{hyperref}
\idline{80}{282}

\begin{document}

\title{Properties of G-band Bright Points  \\ derived from IBIS observations}

\author{S. Criscuoli\inst{1}, D. Del Moro\inst{2}, F. Giorgi\inst{1}, P. Romano\inst{3}, \\
F. Berrilli\inst{2}, I. Ermolli\inst{1}, B. Viticchi\'e\inst{4}, F. Zuccarello\inst{5}}

  \offprints{S. Criscuoli}

\institute{
INAF Osservatorio Astronomico di Roma, Via di Frascati 33, 00040 Monte Porzio Catone, Roma, Italy
\email{serena.criscuoli@oaroma.inaf.it}
\and
Universit\'a degli Studi di Roma Tor Vergata, Italy 
\and
INAF Osservatorio Astrofisico di Catania, Italy
\and
European Space Agency, Noordwijk, The Netherlands
\and
Universit\'a degli Studi di Catania, Italy }

\authorrunning{Criscuoli}

\titlerunning{Properties of G-band  Bright Points  derived from IBIS observations}

\abstract{
We have investigated properties of photospheric Bright Points (BPs) observed in an Active Region during its decay phase and in a quiet Sun region. We have analyzed two sets of photospheric observations taken with IBIS (Interferometric Bidimensioal Spectrometer) at the NSO Dunn Solar Telescope. The first set consists of spectral data acquired in the Fe I 709.0 nm and Ca I 854.2 nm lines and simultaneous broad-band and of G-band observations. 
The second set consists of spectro-polarimetric observations in the Fe I 630.15 nm - 630.25 nm doublet and simultaneous 
white light and G-band observations. 

The relation between BP filling factor and RMS image contrast indicates that, on average, BPs cover up to 3\% of the solar surface outside Active Regions. 
The relation between area and intensity values of the features identified on both data sets suggests that they are 
composed of aggregations of magnetic flux elements. The horizontal velocity values are as high as 2 km/s, thus supporting the scenario of BPs motion contributing to the coronal heating. 
 
}
\maketitle{}

\section{Introduction}

Bright Points (BPs) are small scale features (diameter less than 300 km) corresponding to areas of kilogauss fields. They 
are numerous in active regions and near sunspots, but are ubiquitous in the solar photosphere. The BPs result from processes involving 
the interaction of magnetic fields with convective unstable hot plasma. Their motion can excite Magneto Hydro Dynamic (MHD) waves and trigger 
the occurrence of nano-flares. Both these processes could contribute significantly to the heating of the upper solar atmosphere.

The aim of this study is to investigate the velocity distributions of photospheric BPs, the relation between their filling 
factor and  Root Mean Square (RMS) image contrast, and the 
evolution of different BP characteristics. We present results obtained by analyzing two high spectral, spatial and temporal resolution data sets imaging AR10912 and a quiet Sun region.

\section{Observations}

The data analyzed in this study were acquired with the IBIS \citep{cavallini2006} at DST/NSO on 2006,  October 2$^{nd}$ and November 21$^{st}$. Data sets consist of two time sequences of 
spectral data, spanning about 70 min the first and 41 min the second, respectively. The simultaneous G-band observations cover
 approximately 
the same FOV of the spectral data, but with 4 times smaller pixel scale. The Field of View (FOV) images AR10912 at its decay phase and the quiet Sun, respectively; both regions were observed close to disk centre. 

The first sequence consists of 120 spectral scans of the Fe I 709.0 nm line; each scan contains 1024$\times$1024 pixel images obtained 
at 29 spectral points in the line, with an exposure time of 80 ms and 20 s cadence. This set is complemented with spectral scans at the Ca I 
854.2 nm of the same FOV at 39 spectral points in the line.

The second sequence consists of 50 spectro-polarimetric scans of the Fe I 630.15 nm - 630.25 nm doublet; each scan contains 
256$\times$256 pixel images obtained at 45 spectral points in the lines, with the 6 modulation states I$\pm$V, I$\pm$Q, I$\pm$U; the 
data were stored with 80 ms exposition and 89 s cadence. 

\begin{figure}
\centering{
\includegraphics[width=3.0cm]{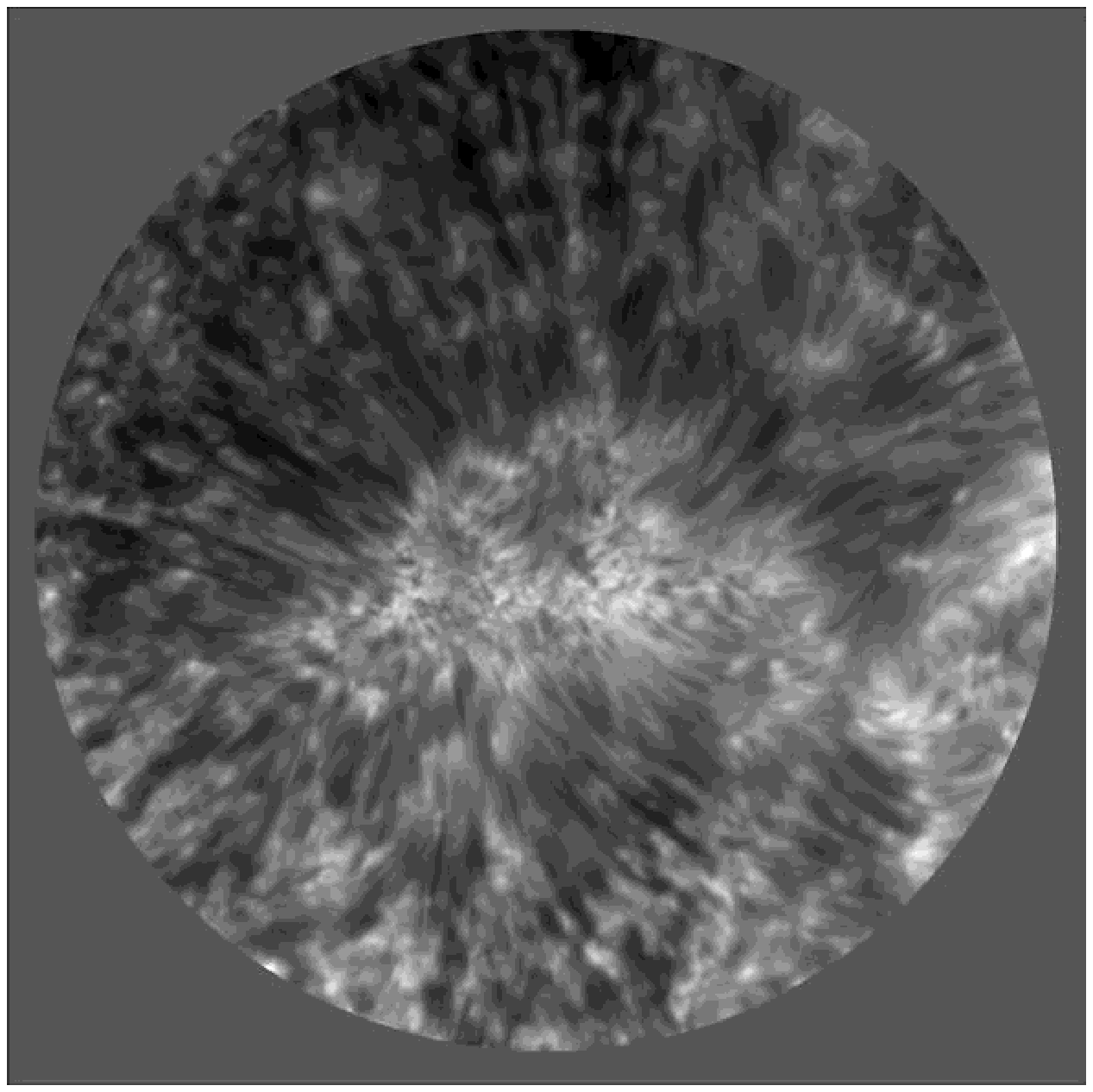}\includegraphics[width=3.0cm]{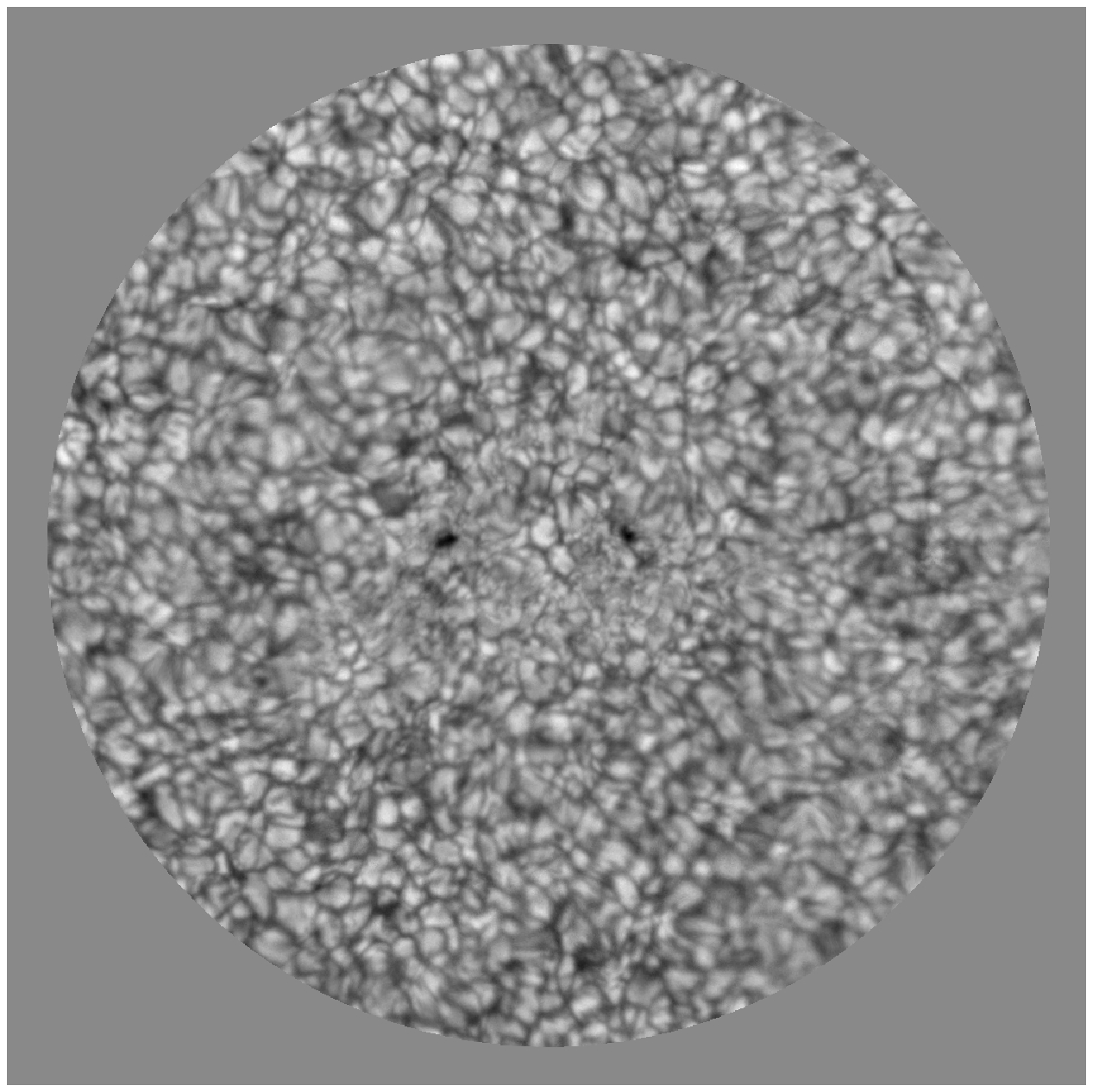}
\includegraphics[width=3.0cm]{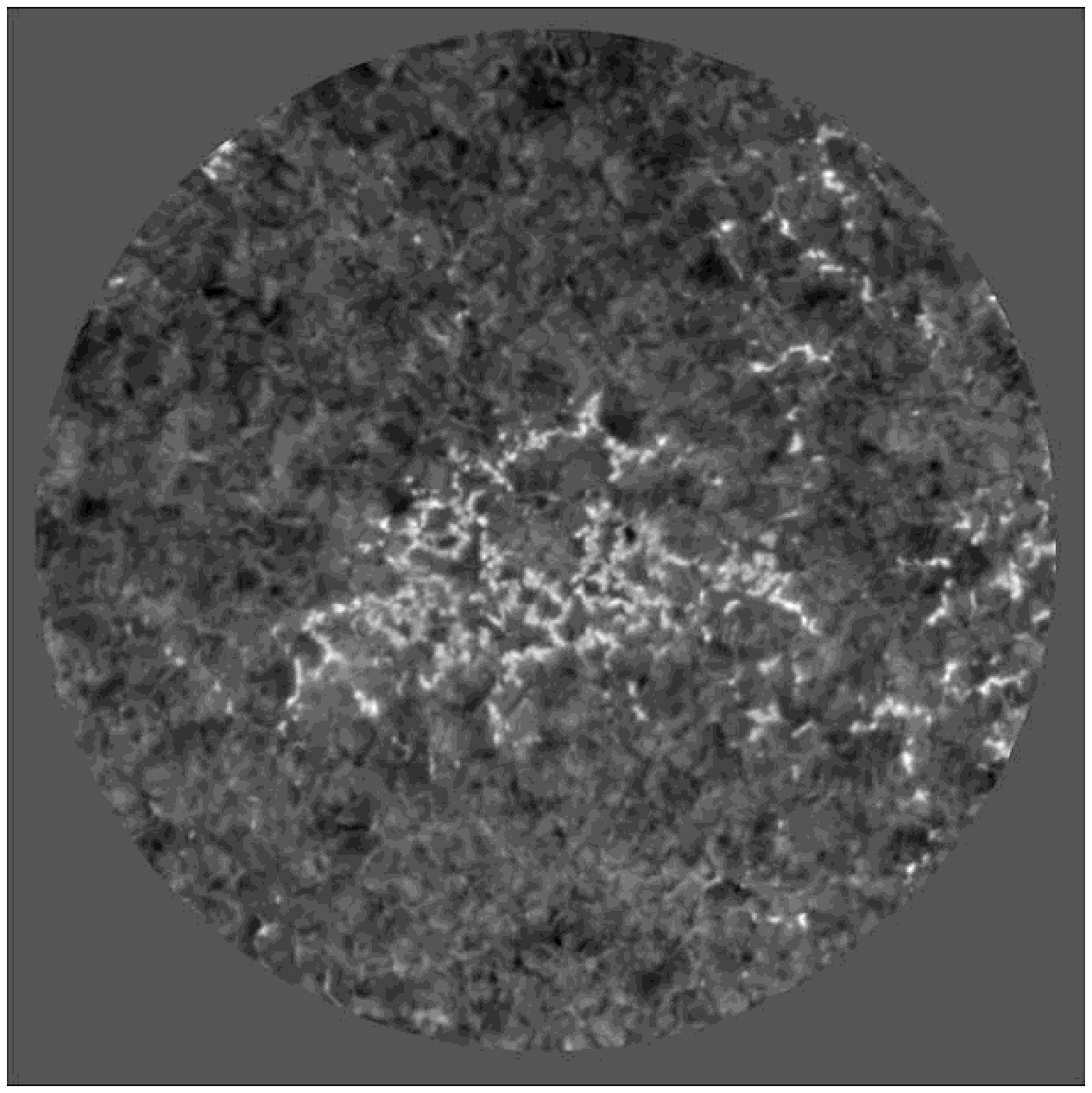}\includegraphics[width=3.0cm]{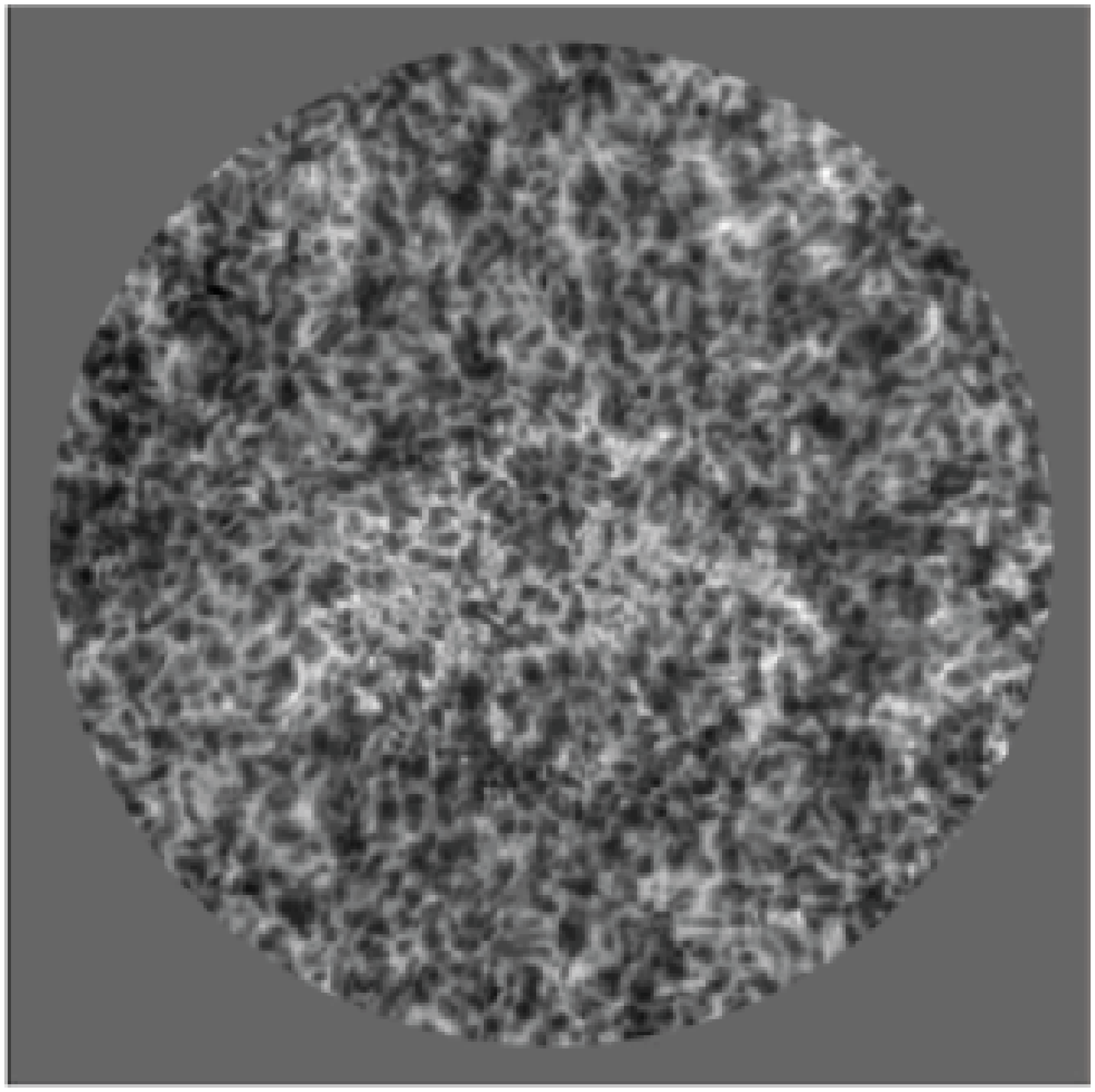}
\includegraphics[width=3.0cm]{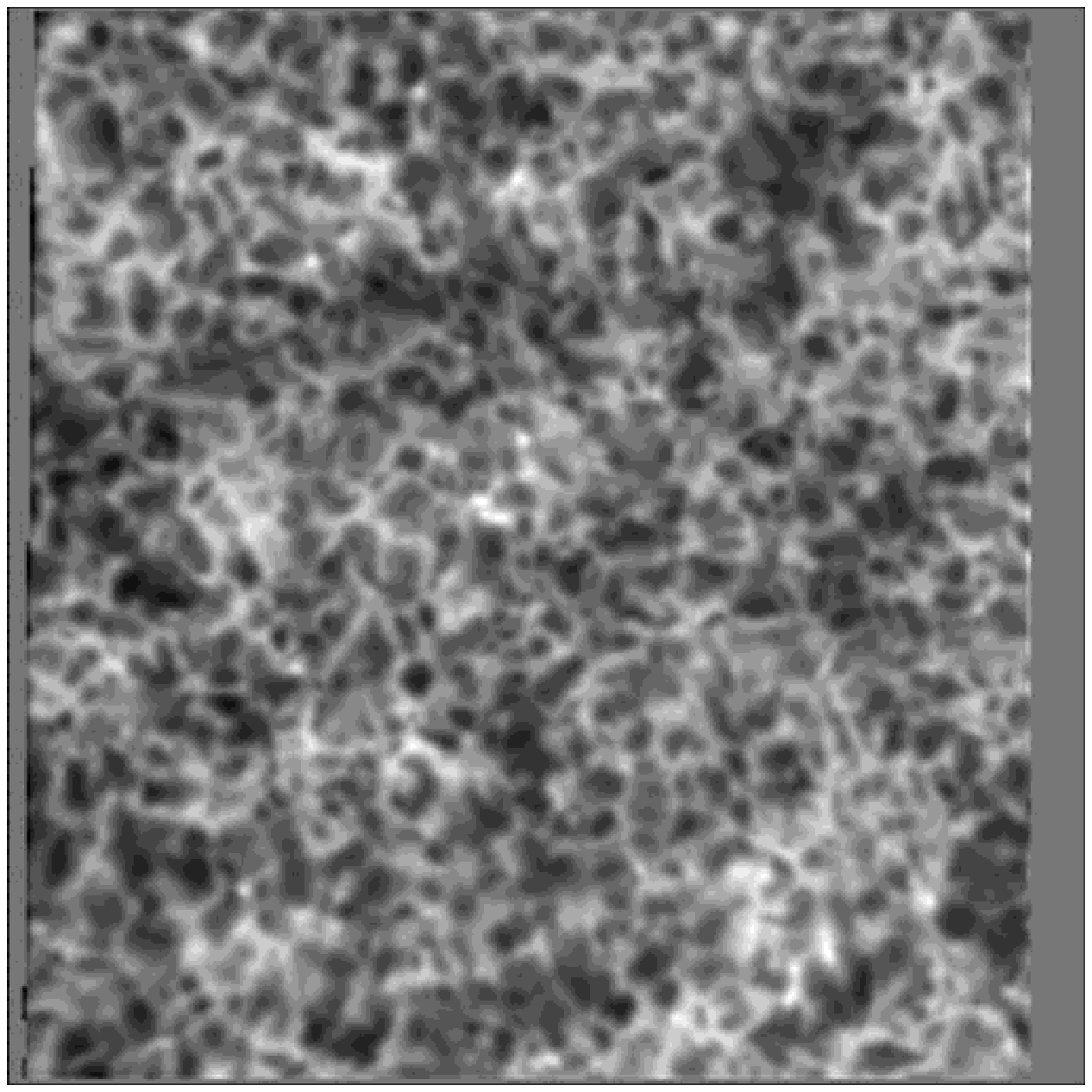}\includegraphics[width=3.0cm]{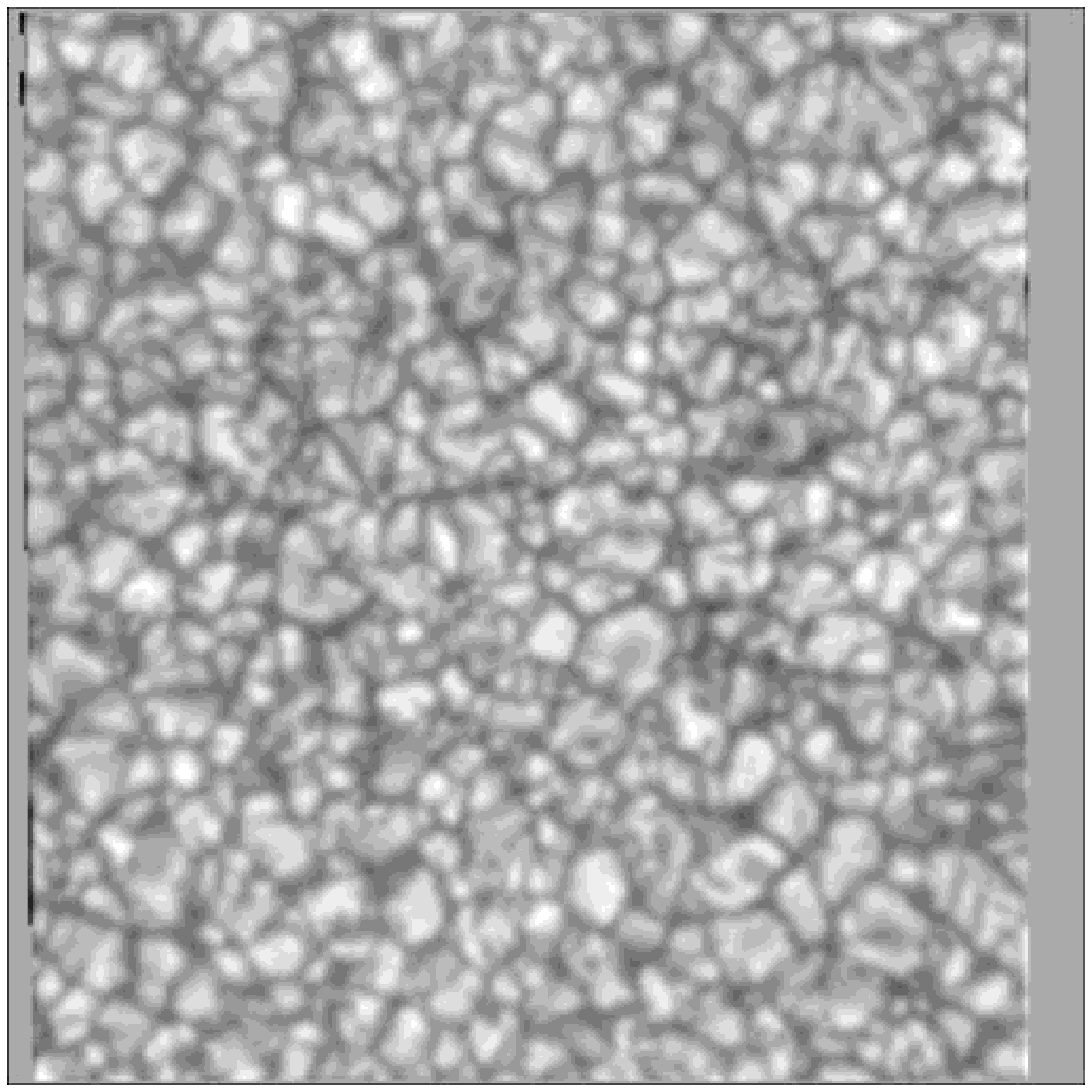}
}
\caption{\footnotesize
Examples of the two data sets imaging the AR (top and middle) and the quiet Sun (bottom) regions analyzed. Top left: Fe I 709.0 nm line core. Top right:  Fe I 709.0 nm line wing. Middle left: 
Ca I 854.2 nm line  
core. Middle right: continuum. The FOV of these data is 80 arcsec diameter; the spatial scale is 0.085 arcsec/pixel. Bottom left: Fe I 630.25 nm 
line core. Bottom right: continuum. The FOV of these data is reduced to 50 arcsec $\times$ 50 arcsec; the spatial scale is 0.17 
arcsec/pixel.}
\label{fig1} 
\end{figure}

Figure \ref{fig1} shows examples of the observations analyzed in this study.  The calibration of data included dark and 
flat-field correction, Multi Frame Blind Deconvolution restoring \citep{vannoort2006} of broadband images, and 
reconstruction of spectral images by re-scaling, rotating, shifting and de-stretching these images to their restored and de-stretched 
broadband counterparts.

\section{Method}
\begin{figure}
\centering{
\includegraphics[width=4.2cm]{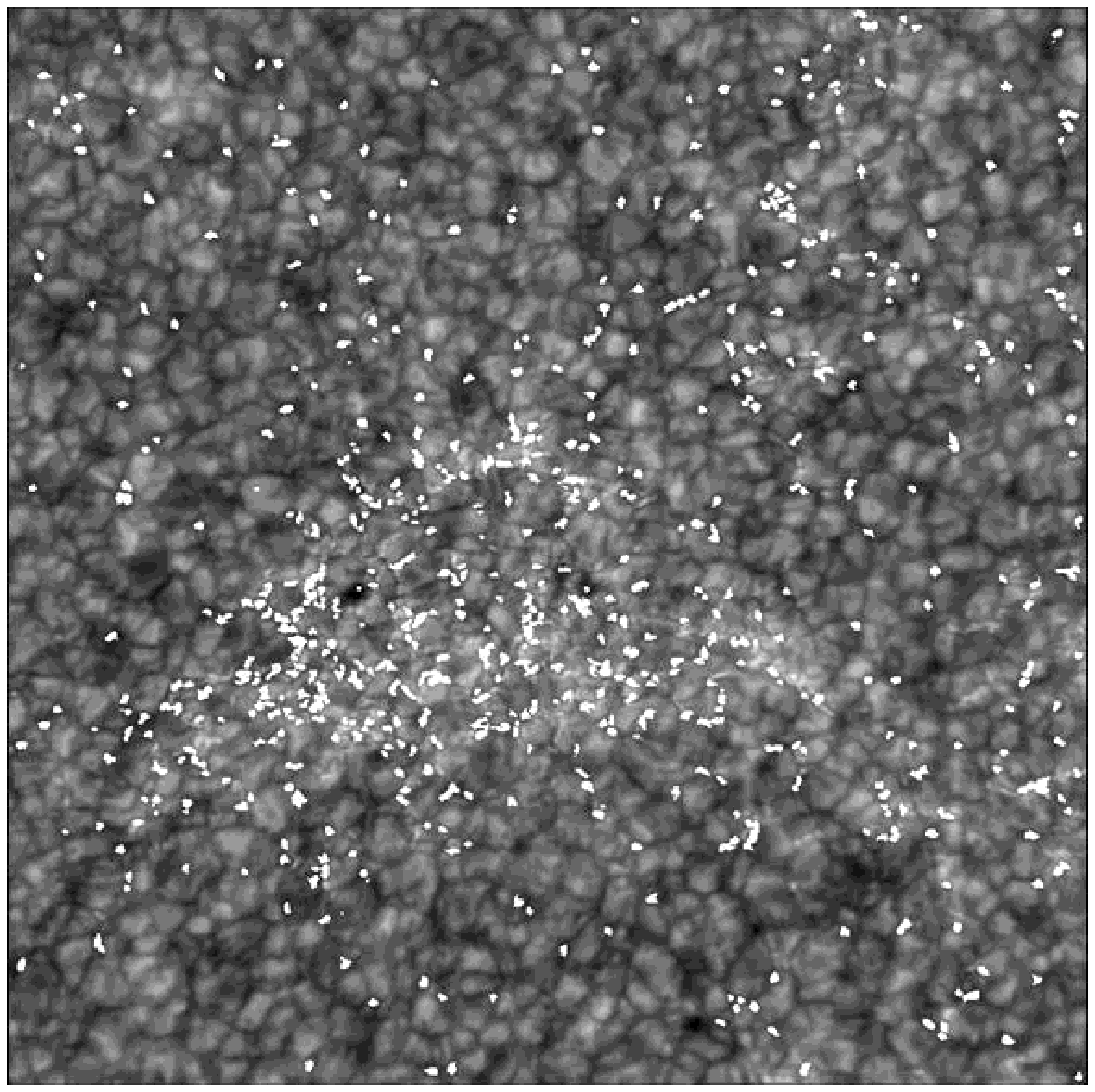}\\
\includegraphics[width=4.2cm]{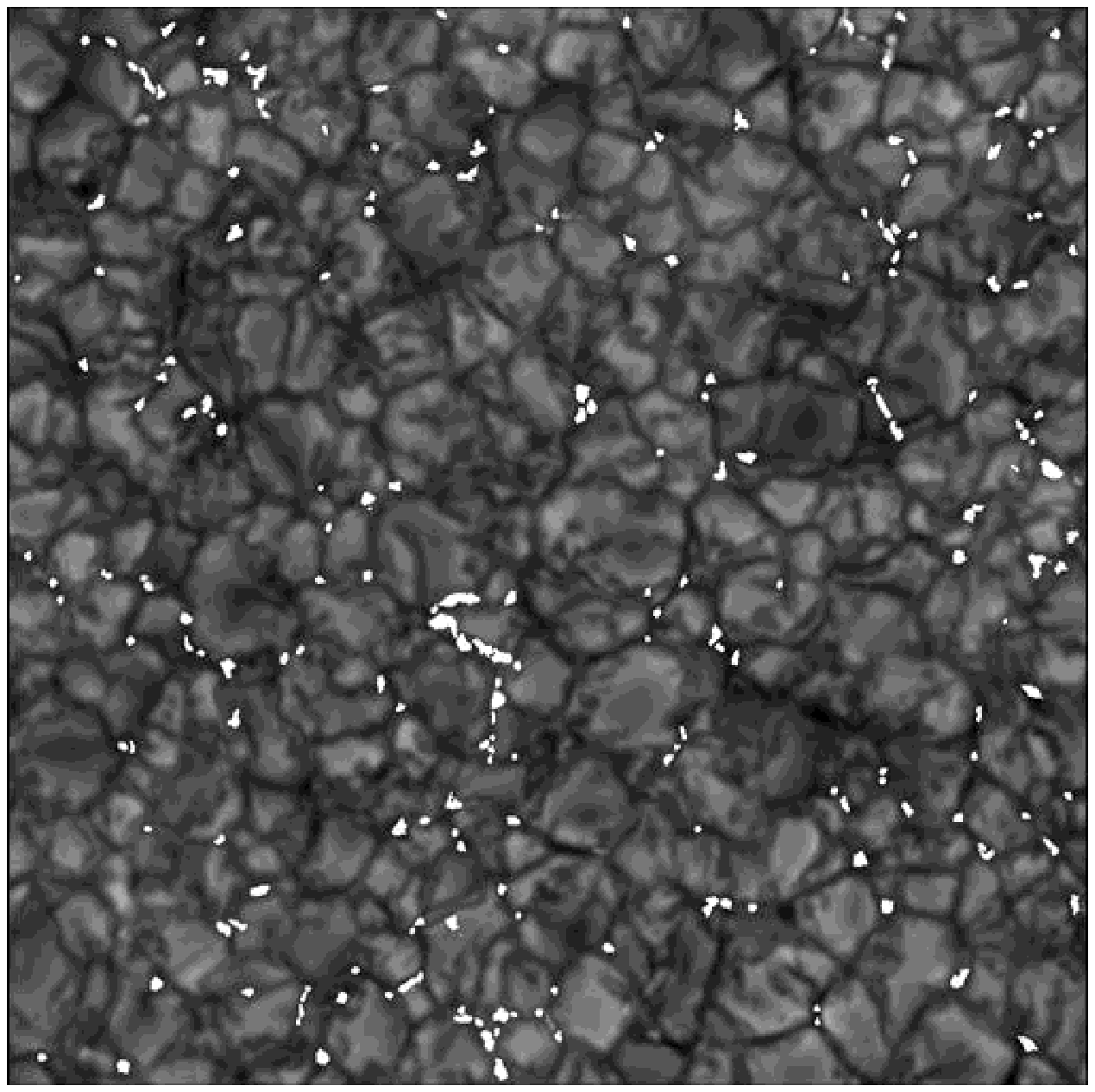}
}
\caption{\footnotesize
Examples of results obtained by applying the BP identification algorithms 
on the G-band images of the two data sets shown in Fig. \ref{fig1}. Top: data imaging AR10912. Bottom: data imaging quiet Sun. 
The
BPs identified by 2-step MLT4 algorithm are overlaid to the segmented image in white.
}
\label{fig2} 
\end{figure}

Small scale BPs were identified  on the G-band images by applying the Multi Level Thresholding 4 (MLT4) algorithm 
\citep{bovelet2007} and a modified version of it (2-step MLT4). Both algorithms produce 
a binary image displaying the BPs identified on the segmented image. To analyze the evolution of the identified features with time,  we then applied 
the Two-level Structure Tracking (TST) algorithm \citep{delmoro2004} on the binary images.
The 2-step MLT4 singled out 5345 and 3120 BPs from all the segmented images of the quiet Sun and AR dataset, respectively. The large number of identified features allows us to 
perform a statistical study 
of the evolution of various BPs properties, and to derive time-averaged values for the analyzed quantities.

\begin{figure} 
\centering{
\includegraphics[width=5.0cm]{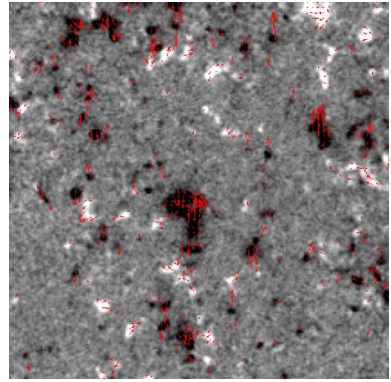}
}
\caption{\footnotesize
Magnetic flux density of the quiet Sun analyzed region, derived from the Fe I 630.25 nm data with the COG method. Red arrows represent the horizontal velocity field derived with the DAVE algorithm. 
}
\label{fig3} 
\end{figure}

To derive an estimate of the longitudinal magnetic flux density, we applied the center-of-gravity method \citep{rees1979} to the left- and right-hand circularly polarized signals of the Fe I  630.25 nm line sequence . 
We also applied the Differential Affine Velocity estimator (DAVE henceforth) algorithm 
\citep{schuck2008} on the magnetic maps computed from the Fe I 630.25 nm data set to 
derive maps of the horizontal velocity fields for the BPs singled out with the image processing. 
Examples of the obtained segmented features, flux density and horizontal velocity maps are shown in  Figs. \ref{fig2} and \ref{fig3}. 

\section{Results and Conclusions}

Figure \ref{fig2} shows  the results obtained by applying the BP identification algorithms 
on the G-band filtergrams co-temporally and co-spatially stored with the spectral data in Fig. \ref{fig1} (bottom panel).  
The
BPs identified by the 2-step MLT4 algorithm are overlaid in white on the segmented G-band image.

We found that the 2-step MLT4
algorithm allows us to identify about 50\% more BPs than those singled out with the MLT4 on the same image. 
We also found that the number of identified BPS increases with the increase of the RMS contrast. Note that, because of stray-light effects and resolution loss due to seeing effects on the data analyzed, the RMS values we measured are $\approx$ 47\% and $\approx$ 70\% in quiet Sun and AR dataset, respectively, lower than the values obtained from numerical simulations \citep{tritschler2006,Wed2009}. From the relation Number of features-RMS contrast deduced from the data analyzed, we extrapolated the number of BPs that would be measured at the RMS contrast values obtained from numerical simulations. In quiet Sun regions we found filling factor values of 2.3\% and 2.6\%  from MLT and 2-step MLT data, respectively; while in AR region we found  4\% and 11.6\%, respectively.  
Note that the filling factor value we found for quiet Sun agrees with the one reported by \citet{sanchez2010} from the analyses of G-band observations taken at better spatial resolution than our observations.

We also analyzed the BP lifetime
and the relation between intensity and area for the identified features. 
The results derived from the two data sets for these quantities show a dependence from the spatial sampling and
seeing conditions of the data sets analyzed. We nevertheless found that, on average, the 
lifetime of features identified in the quiet Sun is $\approx$ 1.5 min, which is in agreement with previous studies \citep{utz2010, deWijn2009}.
In addition, the relation between intensity and area derived from the features identified on both data sets 
suggests that the features analyzed are 
composed by aggregations of unresolved magnetic elements, as obtained with simulations by \citet{criscuoli2009} (see also \citet{viticchie2010} and the discussion therein).

\begin{figure} 
\centering{
\includegraphics[width=6.0cm]{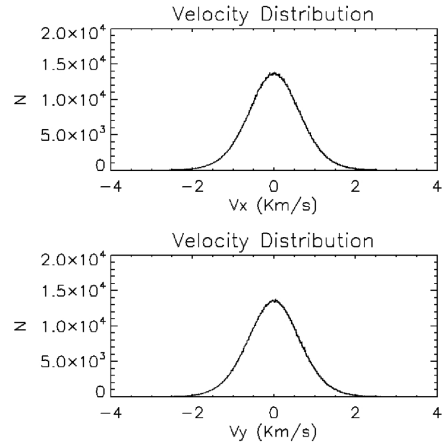}
}
\caption{\footnotesize
Horizontal velocity values derived with DAVE obey Gaussian distributions. 
}
\label{fig4} 
\end{figure}

The horizontal velocity values derived with DAVE on quiet Sun follow a Gaussian distribution, as illustrated in Fig. \ref{fig4}. The velocity values found are as high as 2 km/s, in agreement with those reported by \citet{utz2010} from the analysis of HINODE/SOT G-band observations. Note that this value supports the scenario in which BPs motion could impact the coronal heating.  

\begin{acknowledgements}
The calibration pipeline of IBIS data has been kindly provided by Alexandra Tritschler. 
This study has been partially supported by the INAF, ASI, and MAE,  
within the PRIN-INAF-2007, ASI-ESS and MAE-2007 grants, 
respectively.  
\end{acknowledgements}

\bibliographystyle{aa}

\end{document}